\documentclass[12pt]{article}
\usepackage{amsmath,amssymb}

\begin{document}

\newtheorem{theorem}{Theorem}[section]
\newtheorem{lemma}[theorem]{Lemma}
\newtheorem{defn}{Definition}[theorem]
\renewcommand{\theequation}{\arabic{section}.\arabic{equation}}

\newcommand{\be}{\begin{eqnarray}}
\newcommand{\ee}{\end{eqnarray}}
\newcommand{\bes}{\begin{eqnarray*}}
\newcommand{\ees}{\end{eqnarray*}}
\newcommand{\beqn}{\begin{equation}}
\newcommand{\eeqn}{\end{equation}}

\renewcommand{\theenumi}{(\roman{enumi})}

\newcommand\Pa{Painlev\'e}
\newcommand\Complex{{\Bbb C}}
\newcommand\Real{{\Bbb R}}

\def\sqr#1#2{{\vcenter{\vbox{\hrule height.#2pt
        \hbox{\vrule width.#2pt height#1pt \kern#1pt
           \vrule width.#2pt}
        \hrule height.#2pt}}}}
\def\qed{\begin{flushright}$\sqr84$\end{flushright}}

\title{Nonexistence Results for the Korteweg-deVries and\\
       Kadomtsev-Petviashvili Equations}
\author{Nalini Joshi,\\
         Johannes A. Petersen\thanks{Permanent Address:
Am Sachsenberg 5, 2057 Wentorf bei Hamburg, Germany}\\
and Luke M. Schubert\\
{\it Department of Pure Mathematics}\\
{\it University of Adelaide}\\
{\it Adelaide SA 5005, Australia}\\
}
\maketitle
\begin{center}
Short Title: Nonexistence results for KdV-type equations\
\end{center}

\begin{center}
{\bf Abstract}
\end{center}
We study characteristic Cauchy problems for the
Korteweg-deVries (KdV) equation
$u_t=uu_x+u_{xxx}$, and the Kadomtsev-Petviashvili (KP) equation
$u_{yy}=\bigl(u_{xxx}+uu_x+u_t\bigr)_x$
with holomorphic initial data possessing nonnegative Taylor
coefficients around the origin. For the KdV equation with
initial value $u(0,x)=u_0(x)$,
we show that there is no solution holomorphic
in any neighbourhood of $(t,x)=(0,0)$ in $\Complex^2$
unless $u_0(x)=a_0+a_1x$. This also furnishes a nonexistence result
for a class of $y$-independent solutions of the KP equation. We extend
this to $y$-dependent cases by considering initial
values given at $y=0$, $u(t,x,0)=u_0(x,t)$, $u_y(t,x,0)=u_1(x,t)$,
where the Taylor coefficients of $u_0$ and $u_1$ around $t=0$,
$x=0$ are assumed nonnegative. We prove that there is no
holomorphic solution around the origin in $\Complex^3$
unless $u_0$ and $u_1$ are polynomials of degree 2 or lower.
\begin{center}
AMS Nos: 35Q53, 35B30, 35C10
\end{center}

\section[1]{Introduction}
Completely integrable partial differential equations
such as the Korteweg-deVries (KdV) equation
\[
u_t=uu_x+u_{xxx}
\]
and the Kadomtsev-Petviashvili (KP) equation
\[
u_{yy}=\bigl(u_{xxx}+uu_x+u_t\bigr)_x
\]
are widely believed to have the \Pa\ property \cite{ars:cim,ars:jmp,wtc:jmp},
i.e. all solutions are suspected to be
single-valued around all movable noncharacteristic analytic
singularity manifolds. Although this is a property described in
the complex space of independent variables, very few studies
of the initial value problem in complex space have been
carried out. None to our knowledge have considered
analyticity in $(t,x)\in\Complex^2$.

We carry out such a study
for a restricted
class of holomorphic initial data, as a first step towards
illuminating the \Pa\ property of such equations. In this first
step, we consider real initial data that grow as $\Re(x)\to+\infty$.
As we explain below, these are equivalent to a one-complex-parameter family
of complex initial data.

The initial value problems we study are
\begin{equation}
\left\{\begin{array}{l}
u_t=uu_x+u_{xxx}\\
u(0,x)=u_0(x)
\end{array}
\right.
\label{kdv}
\end{equation}
and
\begin{equation}
\left\{\begin{array}{l}
u_{yy}=\bigl(u_{xxx}+uu_x+u_t\bigr)_x\\
u(t,x,0)=u_0(t,x)\\
u_y(t,x,0)=u_1(t,x)
\end{array}
\right.
\label{kp}
\end{equation}
where $u_0$ and $u_1$ are assumed holomorphic with Taylor
coefficients real and nonnegative. Note that this includes
other cases equivalent to it by a complex changes of variables.
E.g. for the KdV equation, if we take $u(x,t)\mapsto -U(\xi,\tau)$
with $x= i \xi $, $t= -i \tau$, we get
$-i U_{\tau}=-i UU_{\xi}-i U_{\xi\xi\xi}$. The latter is the
same equation, but now the condition on the initial value has changed.
More generally, we can exploit the scaling symmetries of the KdV and
KP equations to transform our hypothesis to allow complex Taylor
coefficients dependent on one parameter. Again for the KdV equation,
under the scaling $u(x,t)=\lambda U(\xi,\tau)$, $\xi=\alpha x$,
$\tau=\beta t$, where $\lambda=\alpha^2$, $\beta=\alpha^3$, our hypothesis
allows initial value $U_0(\xi)$ with $n$-th Taylor coefficients of the form
$\alpha^{-2-n}a_n$ where $a_n$ are real nonnegative, for arbitrary, complex,
nonzero $\alpha$. Note also that by translation invariance, the above
initial value problem can be shifted to the neighbourhood of any complex
point in $x$ or $t$.
A similar equivalent family of initial data is valid for
the KP equation.

We show that unless these data are polynomials in $x$ (of first
degree for the KdV equation and of second degree for the KP
equation), no holomorphic solution exists in any neighbourhood
of the origin in $\Complex^N$, where $N=2$ for the KdV and $N=3$
for the KP equation.
Our main results are
\begin{theorem}
The initial value problem (\ref{kdv}) with initial
data
\begin{equation}
u_0(x) = \sum^{\infty}_{n=0} a_n x^n
\label{an}
\end{equation}
where $a_n\ge 0$ for all $n$, has no solution holomorphic
in any neighbourhood of the origin in $\Complex^2$, unless
$u_0(x)=a_0+a_1x$.
\label{kdv_thm}
\end{theorem}
and
\begin{theorem}
The initial value problem (\ref{kp}) with initial
data
\begin{equation}
u_0(t,x) = \sum^{\infty}_{n=0} c_{0,n}(t) x^n,\quad
u_1(t,x) = \sum^{\infty}_{n=0} c_{1,n}(t) x^n
\label{cjn}
\end{equation}
where $c_{j,n}(t)$ are analytic and have real, nonnegative Taylor
coefficients in $t$ for all $n$, $j=0, 1$, has no solution holomorphic
in any neighbourhood of the origin in ${\Complex}^3$, unless
$u_0$ and $u_1$ are polynomials in $x$ of degree less than or
equal to two.
\label{kp_thm}
\end{theorem}

An illustrative example is given by the initial value
\begin{equation}
u_0(x) = {c\over (a-x)^2}
\label{pole}
\end{equation}
for the KdV equation. If $c>0$ and $a>0$, then
Theorem \ref{kdv_thm} shows that
there is no locally holomorphic solution around the origin.
However, if $c=-12$, it can be easily checked that
this function {\em is} a time-independent
solution of the KdV equation. In the latter case, if $a>0$, all the
Taylor coefficients are real and negative. This example shows that we
cannot enlarge the hypothesis to include purely negative
coefficients in Theorem \ref{kdv_thm}.

The initial data described in
the above theorems are extensions of the usual
initial value problems studied for the KdV and KP equations.
For the KdV equation, inverse scattering theory
\cite{as:siam,ac:cup} shows that there exists a unique solution
for real $u_0$ such that
\begin{equation}
\int_{-\infty}^{\infty} (1+|x|)|u_0(x)| < \infty .
\label{sol}
\end{equation}
In fact, there are well known exact solutions (such
as in example (\ref{pole})) which do not satisfy this condition. But
these solutions have poles on $\Real$.
To our knowledge, there has been no study made of whether the restriction
(\ref{sol}) is necessary for analytic data without poles on $\Real$.  One
motivation for our study is to consider initial values that may be
bounded on the real line but do not necessarily satisfy (\ref{sol})
because of possible growth at infinity. Part of standard PDE theory
is to deduce information such as the admissible order and type of growth
of initial data at infinity. This information is not known for KdV-type
equations.

Existence of solution for the KdV equation for growing initial data
have been studied by Kenig {\it et al} \cite{kpv:97}
in the class of smooth functions on the real
$x$-line and for a half-line in $t$ which in our variables is
$(-\infty, 0]$.
In particular, a classical solution has been shown to exist \cite{kpv:97}
for $t\ge 0$ (in their variables), when $u_0$ is
given by $p(x)+f(x)$ where $p$ is a polynomial of odd degree with
nonnegative coefficients and $f$ is
in the Schwartz class. Our result shows that this classical
solution cannot be holomorphic around the origin except if the degree of
$p$ is unity and $f$ is identically zero.

Finally, we remark here that our method can also be extended to other PDEs,
such as Burgers' equation
\[
u_t=uu_x+u_{xx},
\]
the modified KdV equation
\[
u_t=u^2u_x+u_{xxx},
\]
and
the modified KP equation
\[
u_{yy}=(u_{xxx}+u^2u_x+u_{t})_x.
\]
Similar results hold for these equations with the only change being the
degree of the initial data for which there exists a holomorphic solution.

The proof of Theorem \ref{kdv_thm} and \ref{kp_thm} are given in
Sections 2 and 3 respectively. We also give an alternative proof of
the case of example (\ref{pole}) in Section 2.

\section[2]{Proof of Theorem \ref{kdv_thm}}
Here we prove Theorem \ref{kdv_thm} by studying the formal
solution
\begin{equation}
u(x,t)=\sum_{n=0}^{\infty}u_n(x)t^n,
\label{formal}
\end{equation}
in particular, the recursion relation satisfied by its coefficients.
We divide the proof into three cases. The first is when
$u_0(x)$ is polynomial. We show that if $\deg(u_0)\ge 2$ then $u_n$ must grow
factorially in
$n$. Then we show that the proof extends simply to the case of
nonpolynomial $u_0(x)$. Finally we show that there does exist a holomorphic
solution, in a neighbourhood of the origin in $\Complex^2$, when $u_0(x)$ is a
linear function of $x$.

Note that the coefficients $u_n(x)$ satisfy
\begin{equation}
(n+1)u_{n+1}=\sum^{n}_{k=0}u_ku'_{n-k}+u'''_n
\label{rec}
\end{equation}
\subsection{The Polynomial Case}
Here we consider the case of polynomial $u_0(x)$ of degree $d_0\ge 2$.
Clearly, $u_n$ must be polynomial
if $u_0$ is. Let $d_n$ be its degree and write
\[u_n = \sum_{k=0}^{d_n}c_{n,k}x^{d_n-k}\]
Assume that $c_{0,0}>\epsilon$ for some $0<\epsilon<1$.
\begin{lemma}
The degree $d_n$ and coefficient $c_{n,0}$ of the largest
degree term in $u_n$ satisfy
\be
d_n&=&(n+1)d_0-n\\
c_{n,0}&>&\epsilon^{n+1},
\ee
\end{lemma}
\noindent{\bf Proof:} The proof is by induction.
First note that the third derivative term
on the right side of (\ref{rec}) acts to decrease degree,
whereas the convolution term increases it. We have at $n=1$
\be
d_1 &=& 2d_0 - 1\\
c_{1,0} &=& d_0c_{0,0}^2>2\epsilon^2 > \epsilon^2
\ee
Now suppose that the results hold for $1, \ldots, n$. Consider
the $(n+1)$-st case. Then the maximal degree of the convolution
term is given by
\[(k+1)d_0-k-1+(n-k+1)d_0-(n-k) = (n+2)d_0-(n+1)\]
for $0\le k\le n$, which proves the result for $d_{n}$ for all $n\ge
1$.
The defining equation for the coefficient $c_{n,0}$ is
\bes
(n+1)c_{n+1,0}&=&\sum_{k=0}^nd_kc_{k,0}c_{n-k,0}\\
              &>&\sum_{k=0}^nc_{k,0}c_{n-k,0}\\
              &>&(n+1)\epsilon^{n+2}
\ees
by the induction hypothesis. Hence the result holds for $c_n$
for all $n\ge 1$. \qed

We now focus on the third derivative term in Eqn(\ref{rec}) to
show divergence. Its contribution to lower-degree
terms can be estimated as follows. First we identify the
degree of its contribution by using
\be
d_{3n}-3 &=& d_{3n+1}-(d_0+2)
\label{red1}\\
\vdots& &\nonumber\\
d_{3n+m}-l_m-3 &=&d_{3n+m+1}-l_{m+1}
\label{red2}
\ee
where $l_{m+1}=l_m+d_0+2$ which implies $l_m=m(d_0+2)$.
Therefore, we get:
\begin{lemma}
For $3(m+1)\le d_{3n}$,
$c_{3n+m+1,l_{m+1}}$ is lowerbounded by
\[
c_{3n+m+1,l_{m+1}} > {(d_{3n+m}-l_m+3m)!(3n)!\over
(d_{3n+m}-l_m-3)!(3n+m+1)!}c_{3n,0}
\]
\label{lbc}
\end{lemma}
\noindent{\bf Proof:} From the third derivative term,
by using nonnegativity, we get
\[
c_{3n+m+1,l_{m+1}} >
{(d_{3n+m}-l_m)(d_{3n+m}-l_m-1)(d_{3n+m}-l_m-2)\over
(3n+m+1)}c_{3n+m,l_m}.
\]
The desired result follows from a recursive use of this
inequality and the relations (\ref{red1}--\ref{red2}) which give
\[
{(d_{3n+m-i}-l_{m-i})!\over (d_{3n+m-i}-l_{m-i}-3)!}
   = {(d_{3n+m-i+1}-l_{m-i+1}+3)!\over (d_{3n+m-i+1}-l_{m-i+1})!}
\]
for $0\le i\le m$.

Now choose $m=n-1$ (which satisfies $3(m+1)\le d_{3n}$
because $d_0\ge 2$).
Then we get
\[
c_{4n,l_{n}} > {\bigl(3n(d_0-1)+d_0\bigr)!(3n)!\over
\bigl(3n(d_0-2)+d_0\bigr)!(4n)!}\epsilon^{3n+1}.
\]
Since this grows factorially with $n$, for $d_0\ge 2$, we get divergence
for the formal series (\ref{formal}). \qed

\subsection{The Nonpolynomial Case}
Now suppose $u_0$ is nonpolynomial. Then $\{u_n\}$ are no longer polynomial, so
we write
\[u_n = \sum_{p=0}^{\infty}b_{n,p}x^{p}.\]
Assume, as before, that $b_{0,d_0}>\epsilon$ for some $0<\epsilon<1$ and
$d_0\ge
2$. (Note that $d_0$ is no longer the degree of $u_0$.) Then from the
recursion relation (\ref{rec}) we again get that, for all $n\ge 1$ and
$d_n:=n(d_0-1)+d_0$, the coefficients
$b_{n, d_n}$ have a lower bound
\[b_{n,d_n}>\epsilon^{n+1}.\]
The remainder of the argument for the polynomial case now follows
to give a factorially growing lowerbound for
$b_{4n,q_n}$, where $q_n:= d_{4n}-l_n = 3n(d_0-2)+d_0$.

\subsection{The Linear Case}
Here we consider the case $u_0(x)=a_0+a_1x$. In this case, there exists an
exact solution:
\[u(x, t)={a_0+a_1x\over 1-a_1t}.\]
This solution is clearly holomorphic in the polydisk
\[\left\{(x, t)| x\in\Complex, |t|< 1/|a_1|\right\}.\]
Note that this result holds even if $a_0$, $a_1$ are not nonnegative.

\subsection{The Case of an Isolated Double Pole}
Consider the example (\ref{pole}) as initial datum with $a=1$.
(We can use the translational invariance of the KdV equation
to assume this without loss of generality.) We give an alternate,
simple, proof of nonexistence for this case here.

We claim that the coefficients $u_n(t)$ of \ref{formal}
can be written as
\[
u_n(t) = {a_n\over (1-x)^{3n+2}}.
\]
The inductive proof follows from the substitution of this
form into the right side of (\ref{rec}):
\bes
& &\sum^{n}_{k=0}u_ku'_{n-k}+u'''_n\\
\quad\quad&=& \sum^{n}_{k=0}
{\bigl(3(n-k)+2\bigr)a_ka_{n-k}\over (1-x)^{3n+5}}\\
\quad\quad& &\quad + {(3n+2)(3n+3)(3n+4)a_n\over ((1-x)^{3n+5}}
\ees

So the recursion relation satisfied
by $a_n$ is
\bes
(n+1)a_{n+1}&=&\sum^{n}_{k=0}\bigl(3(n-k)+2\bigr)a_ka_{n-k}\\
\quad\quad& &\quad + (3n+2)(3n+3)(3n+4)a_n\\
&\ge&(3n+2)(3n+3)(3n+4)a_n
\ees
Hence the $a_n$ are lower bounded by
\[
a_n\ge {(3n+1)! c\over n!}.
\]
Clearly these give rise to a divergent series (\ref{formal}).

\section{Proof of Theorem \ref{kp_thm}}
For the KP equation, we consider the formal solution
\begin{equation} \label{E:KPnew}
u(t,x,y) = \sum_{j=0}^{\infty} u_j(t,x) y^j,
\end{equation}
whose coefficients satisfy the recurrence relation:
\begin{equation} \label{E:KPrr}
(n+2)(n+1) u_{n+2} = u_{n,4x} + \sum_{j=0}^n (u_j u_{n-j,x})_x
+u_{n,tx}
\end{equation}
for $n \ge 2$. We follow a similar argument to that of the previous
section and give the details of the proof for polynomial $u_0$, $u_1$
of respective degree $d_0\ge 3$, $d_1\ge 3$ in
$x$. The argument for the nonpolynomial $u_0$ or $u_1$ is the same
under the assumption that $d_0$ and $d_1$ are no longer their
degrees but the degrees of some terms in their Taylor expansion
that we track in the recursive
estimate of coefficients.

Suppose that $u_0$ and $u_1$ are both non-negative
polynomials in $x$ (that is, polynomials with all coefficients
non-negative for all $t$),
and are analytic functions of $t$, so that
\begin{equation}
 u_j(t,x) = \sum_{k=0}^{d_j} c_{j,k}(t) x^{d_j-k}
\label{uj}
\end{equation}
for $j=0, 1$, where $c_{j,k}(t)$ are analytic functions
of $t$ with all Taylor coefficients non-negative.

Then $u_n$, for $n \ge 2$, will
likewise be a non-negative polynomial in $x$ for all $t$.
We assume that $u_j$ for $j\ge 2$ have the form given by (\ref{uj}).
Since we have such non-negativity, we can generally ignore the term
$u_{tx}$ in (\ref{kp}) and (\ref{E:KPrr}) for purposes of
calculating lower bounds.

The first question is, what degree are
the polynomials $u_j$?
The degree $d_j$ follows different patterns (with respect to $j$)
depending on whether $3d_0$ is greater than or less than
$2d_1+2$. We consider these cases separately, and also describe
what happens when $u_0$ or $u_1$ is identically zero.

\begin{lemma} \label{L:KPdeg}
\begin{enumerate}
\item If $3d_0 \ge 2d_1 + 2$, then for all $n \ge 0$,
   $d_{2n} = n(d_0-2) + d_0$ and $d_{2n+1} = n(d_0-2) + d_1$.
\item If $3d_0 < 2d_1 + 2$, then for all $n \ge 0$,
   $d_{3n} = n(d_1-2) + d_0, d_{3n+1} = n(d_1-2)+d_1$ and
   $d_{3n+2} = n(d_1-2) + 2d_0 -2$.
\item If $u_1 = 0$, then for all $n \ge 0$, $d_{2n} = n(d_0-2)+d_0$
   and $u_{2n+1} = 0$.
\item If $u_0 = 0$, then $u_2=0$ and for all $n \ge 1$,
   $d_{3n} \le n(d_1-2)+1$,
   $d_{3n+1} = n(d_1-2)+d_1$ and $d_{3n+2} \le n(d_1-2)$.
\end{enumerate}
\end{lemma}

\noindent{\bf Proof:}
In all cases, we obtain from (\ref{E:KPrr}) the following relation:
\begin{equation} \label{E:KPdeg}
d_k = \max \{ d_0 + d_{k-2} -2, \dots , d_{k-2} + d_0 -2, d_{k-2} -1 \}
\end{equation}
or equivalently:
\[ d_k = \max_{j=0, \dots, k-2} \{ d_j + d_{k-2-j}-2, d_{k-2}-1 \} \]
where the last term is only present if $c_{k-2,0}(t)$ is not constant,
and is not necessary (and not mentioned) below, with the exception of
(iv).

(i) The claim is true for $k=0$ by definition of $d_0$ and $d_1$.
Assume that it is true for all $n \le k$, for some $k$.
Then we use (\ref{E:KPdeg}) as follows:
\begin{eqnarray*}
d_{2k+2} & = & \max \{ d_{2j}+d_{2k-2j} -2, d_{2j+1}+d_{2k-2j-1}-2 \}
		\\
         & = & \max \{ k(d_0 -2) +2d_0 -2, (k-1)(d_0-2) +2d_1-2 \}
		\\
         & = & k(d_0-2)+2d_0-2 \\
d_{2k+3} & = & \max \{ d_{2j}+d_{2k-2j+1} -2, d_{2j+1}+d_{2k-2j}-2 \} \\
         & = & \max \{ k(d_0-2)+d_0+d_1-2, k(d_0-2)+d_1+d_0-2 \} \\
	 & = & k(d_0-2) + d_0-2+d_1.
\end{eqnarray*}
In the first series of equations, we use the assumption that $3d_0 \ge 2d_1
+2$; also, $j$ ranges from $0$ to $k$.

(ii) For $k=0$, we need only check that $d_2=2d_0-2$; but this follows since
$2u_2 = u_{0,4x} + u_{0,tx} + (u_0 u_{0,x})_x$ (by (\ref{E:KPrr})).
Assume that the claim is true for all $n \le k$, for some $k$.
Then we again use (\ref{E:KPdeg}) as follows:
\begin{eqnarray*}
d_{3k+3} & = & \max \{ d_{3j}+d_{3(k-j)+1}-2,d_{3j+1}+d_{3(k-j)}-2,d_{3j+2} +
		d_{3(k-j-1)+2}-2 \} \\
	 & = & \max \{ k(d_1-2)+d_0+d_1-2, k(d_1-2)+d_1+d_0-2, \\
	 & & (k-1)(d_1-2) +4d_0-6 \} \\
	 & = & (k+1)(d_1-2)+d_0 \\
d_{3k+4} & = & \max \{ d_{3j}+d_{3(k-j)+2}-2,d_{3j+1}+d_{3(k-j)+1}-2,d_{3j+2} +
		d_{3(k-j)}-2 \} \\
	 & = & \max \{ k(d_1-2)+3d_0-4, k(d_1-2)+2d_1-2, k(d_1-2)
		+3d_0-4 \} \\
	 & = & (k+1)(d_1-2)+d_1 \\
d_{3k+5} & = & \max \{ d_{3j}+d_{3(k-j+1)}-2,d_{3j+1}+d_{3(k-j)+2}-2,d_{3j+2} +
		d_{3(k-j-1)+1}-2 \} \\
	 & = & \max \{ (k+1)(d_1-2)+2d_0-2, k(d_1-2)+d_1+2d_0-4, \\
	 & & k(d_1-2)
		+2d_0+d_1-4 \} \\
	 & = & (k+1)(d_1-2)+2d_0-2
\end{eqnarray*}
Throughout, $j$ varies between $0$ and $k-1$, and we use the assumption
that $3d_0 < 2d_1+2$ in the first two series of equations.

Thus, by induction, the claim is true for all integers $n \ge 0$.

(iii) The claim is trivially true for $n=0$. Suppose it is true for all
$n \le k$.
Then by (\ref{E:KPrr}),
$(2n+3)(2n+1)u_{2n+3} = \sum_{j=0}^{2n+1} (u_j u_{2n+1-j,x})_x$;
but each of the terms in the sum is identically zero (either $j$ is odd,
or $2n+1-j$ is odd).
Also by (\ref{E:KPrr}), we have that
\[ d_{2n+2} = \max \{ d_0 + d_{2n}-2, d_2 + d_{2n-2}-2, \dots, 2d_n-2 \}. \]
But each of these numbers is just $(n+1)(d_0-2)+d_0$.
Thus, by induction, the claim is true for all $k \ge 0$.

(iv) In this case, the degree can take values below the maximum (and in some
cases, $u_{3n+2}$ can be $0$) if, for example, $c_{1,0}'(t)$ is identically
zero.

Since $u_0=0$, we have that $u_2=0$.
Assume that the claim is true for all $n \le k$, for some $k$.
Then we again use (\ref{E:KPdeg}) as follows:
\begin{eqnarray*}
d_{3k+3} & \le & \max \{ d_{3j}+d_{3(k-j)+1}-2,d_{3j+1}+d_{3(k-j)}-2,d_{3j+2} +
		d_{3(k-j-1)+2}-2, \\
	 & &  d_{3k+1}-1 \} \\
	 & \le & \max \{ k(d_1-2)+d_1-1, k(d_1-2)+d_1-1, (k-1)(d_1-2) -2, \\
	 & & k(d_1-2)+d_1-1 \} \\
	 & = & (k+1)(d_1-2)+1
\ees
\bes
d_{3k+4} & = & \max \{ d_{3j}+d_{3(k-j)+2}-2,d_{3j+1}+d_{3(k-j)+1}-2,d_{3j+2} +
		d_{3(k-j)}-2, \\
	 & & d_{3k+2}-1  \} \\
	 & = & \max \{ k(d_1-2)-1, k(d_1-2)+2d_1-2, k(d_1-2)-1, k(d_1-2)-1\} \\
	 & = & (k+1)(d_1-2)+d_1
\ees
\bes
d_{3k+5} & \le & \max \{ d_{3j}+d_{3(k-j+1)}-2,d_{3j+1}+d_{3(k-j)+2}-2,
	 d_{3j+2} + d_{3(k-j-1)+1}-2 \\
	 & & d_{3k+3}-1 \} \\
	 & = & \max \{ (k+1)(d_1-2), k(d_1-2)+d_1-2, \\
	 & & k(d_1-2)-1, (k+1)(d_1-2) \} \\
	 & = & (k+1)(d_1-2)
\end{eqnarray*}
Throughout, $j$ varies between $0$ and $k-1$.
\qed

{}From the above lemma, we see that the degree of the polynomial $u_n$ in
$x$ will only grow (as $n$ increases) if either $d_0 \ge 3$ or $d_1 \ge 3$.
If so, and if $c_{0,0}(t)$ or $c_{1,0}(t)$ (where they exist)
have a positive lower bound, the leading coefficients of certain of the
terms $u_n$ grow exponentially, as follows.

\begin{lemma} \label{L:KPcoeff}
\begin{enumerate}
\item If $3d_0 \ge 2d_1+2$, $d_0 \ge 3$ and for all $t>0$,
	$c_{0,0}(t)> \epsilon$
	and $c_{1,0}(t) > \epsilon$, then for all $n \ge 0$,
	$c_{2n,0}(t) > \epsilon^{n+1}$ and $c_{2n+1,0}(t) > \epsilon^{n+1}$.
\item If $3d_0 < 2d_1+2$, $d_1 \ge 3$ and for all $t>0$, $c_{1,0}(t) >
	\epsilon$, then for all $n \ge 0$, $c_{3n+1,0}(t) > \epsilon^{n+1}$.
\item If $u_1=0$, $d_0 \ge 3$ and for all $t>0$, $c_{0,0}(t)> \epsilon$, then
	for all $n \ge 0$, $c_{2n,0}(t) > \epsilon^{n+1}$.
\item If $u_0=0$, $d_1 \ge 3$ and for all $t>0$, $c_{1,0}(t) > \epsilon$, then
	for all $n \ge 0$, $c_{3n+1,0}(t) > \epsilon^{n+1}$.
\end{enumerate}
\end{lemma}

\noindent{\bf Proof:}
In all cases, we begin by deducing the following relation from (\ref{E:KPrr}):
\begin{equation} \label{E:KPcoeff}
k(k-1)c_{k-1,0}(t) \ge (d_k+1) \sum_{j=0}^{k-2} d_j b_{j,k} c_{j,0}(t)
	c_{k-2-j,0}(t).
\end{equation}
Here $b_{j,k}$ is simply a constant, either $0$ or $1$ depending on which
terms of the form $(u_j u_{k-2-j,x})_x$ contribute to the highest-order
term $x^{d_k}$.

(i) Assume that the claim is true for all $n \le k$.
Using all our assumptions, we have that
\begin{eqnarray*}
(2k+2)(2k+1) c_{2k+2,0}(t) & > & (d_{2k+2}+1) \epsilon^{k+2}
	\sum_{j=0}^{k} d_{2j} \\
	& = & \epsilon^{k+2} ((k+1)(d_0-2)+d_0+1) \\
	& & (k+1)\left( \frac{1}{2}k(d_0-2) +d_0 \right) \\
	& \ge & (2k+2)(2k+1) \epsilon^{k+2} \ \rm{since} \ d_0 \ge 3
\ees
\bes
(2k+3)(2k+2) c_{2k+3,0}(t) & > & (d_{2k+3}+1) \epsilon^{k+2}
	\sum_{j=0}^{2k} d_j \\
	& = & \epsilon^{k+2} ((k+1)(d_0-2)+d_1+1) \\
	& & (k+1)(k(d_0-2)+d_0+d_1) \\
	& \ge & (2k+3)(2k+2) \epsilon^{k+2}.
\end{eqnarray*}
Note that all terms contribute in the second case, but only odd terms in
the first.

(ii) Assume that the claim is true for all $n \le k$.
Then we have that
\begin{eqnarray*}
%(3k+3)(3k+2) c_{3k+3,0}(t) & \ge & (d_{3k+3}+1) \epsilon^{2k+3}
%	\sum_{j=0}^{k} d_{3j} + d_{3j+1} \\
%	& = & \epsilon^{2k+3} ((k+1)(d_1-2)+d_0+1)(k+1)(k(d_1-2)+d_0+d_1) \\
%	& \ge & (3k+3)(3k+2) \epsilon^{2k+3} \ \rm{since} \ d_1 \ge 3 \\
		% CHECK the last step ... in all these cases.
		% In fact, the last inequality only holds, here and below,
		% for large enough k (and/or for d_1 \ge 4).
(3k+4)(3k+3) c_{3k+4,0}(t) & > & (d_{3k+4}+1) \epsilon^{k+2}
	\sum_{j=0}^{k} d_{3j+1} \\
	& = & \epsilon^{k+2} ((k+1)(d_1-2)+d_1+1) \\
	& & (k+1)\left( \frac{1}{2}k(d_1-2) +d_1 \right) \\
	& \ge & (3k+4)(3k+3) \epsilon^{k+2} \ \rm{if} \ d_1 \ge 4 \ \rm{or}
	\ k \ge 6. \\
%(3k+5)(3k+4) c_{3k+5,0}(t) & \ge & (d_{3k+5}+1) \epsilon^{2k+4}
%	\sum_{j=0}^{3k+3} d_{j} \\
%	& \ge & \epsilon^{2k+4} ((k+1)(d_1-2)+2d_0-1)(k+1) \\
%	& & (\frac{3}{2}k(d_1-2)+3d_0+d_1 -2) \\
%	& \ge & (3k+5)(3k+4) \epsilon^{2k+4} \ \rm{if} \ d_1 \ge 3
\end{eqnarray*}
The cases $d_1 =3$ and $k=1, \dots, 5$ can be calculated explicitly
(and fully, using exact values) to show that in these cases also,
$c_{3k+4,0}(t)$ is greater than $\epsilon^{k+2}$.

So the claim is true for all integers $n \ge 0$.

(iii) Assume that the claim is true for all $n \le k$.
Then we have that
\begin{eqnarray*}
(2k+2)(2k+1) c_{2k+2,0}(t) & > & (d_{2k+2}+1) \epsilon^{k+2}
	\sum_{j=0}^{k} d_{2j} \\
	& = & \epsilon^{k+2} ((k+1)(d_0-2)+d_0+1) \\
	& &  (k+1)\left( \frac{1}{2}k(d_0-2)+d_0 \right) \\
	& \ge & (2k+2)(2k+1) \epsilon^{k+2} \ \rm{if} \ d_0 \ge 3
\end{eqnarray*}

(iv) %is a whole new ball game.
The proof here is identical to that in (ii), since the degree of $u_{3k+1}$
is the same in both cases (and since only terms of the form $c_{3j+1,0}(t)$
contribute to $c_{3k+1}(t)$ in both cases). \qed

We now follow the method used for the KdV equation.
%Set
%$l_{2m}:= m(d_0 +2)$; we

\begin{lemma}
\begin{enumerate}
\item If $3d_0 \ge 2d_1 +2$, $d_0 \ge 3$ and $c_{0,0}(t) > \epsilon$ etc.,
then for all positive integers $q$,
\[c_{10q, q(d_0+2)} >
\frac{(4q(d_0-2)+d_0)! (8q)!}{(4q(d_0-3)+d_0)! (10q)!} \epsilon^{4q+1}.\]
\item If $3d_0 < 2d_1 +2$, $d_1 \ge 3$ and $c_{0,0}(t) > \epsilon$ etc.,
then for all positive integers $q$,
\[c_{42q+1, 2q(d_1+4)} > \frac{(12q(d_1-2)+d_1)! (36q+1)!}{12q(d_1-3)+d_1)!
(42q+1)!} \epsilon^{12q+1}.\]
\item If $u_1 = 0$, $d_0 \ge 3$ and $c_{0,0}(t) > \epsilon$ etc.,
then for all positive integers $q$,
\[c_{10q, q(d_0+2)} >
\frac{(4q(d_0-2)+d_0)! (8q)!}{(4q(d_0-3)+d_0)! (10q)!} \epsilon^{2q+1}.\]
\item If $u_0 = 0$, $d_1 \ge 3$ and $c_{1,0}(t) > \epsilon$ etc.,
then for all positive integers $q$,
\[c_{42q+1, 2q(d_1+4)} > \frac{(12q(d_1-2)+d_1)! (36q+1)!}{12q(d_1-3)+d_1)!
(42q+1)!} \epsilon^{12q+1}.\]
\end{enumerate}
\end{lemma}

\noindent{\bf Proof:}
{}From (\ref{E:KPrr}), we see due to the $u_{n-2,xxxx}$ term and the
non-negativity of the coefficients that the coefficient of $x_{d_p}$ in
$u_p$, $c_{p,0}(t)$, adds to the coefficient of $x^{d_p-4}$ in $u_{p+2}$,
$c_{p+2,d_{p+2}-d_p+4}$. That is,
\[ c_{p+2,d_{p+2}-d_p+4} > \frac{d_p(d_p-1)(d_p-2)(d_p-3)}{(p+2)(p+1)}
c_{p,0}. \]
Similarly,
\[ c_{p+4,d_{p+4}-d_p+8} > \frac{(d_p-4)(d_p-5)(d_p-6)(d_p-7)}{(p+4)(p+3)}
c_{p+2,d_{p+2}-d_p+4}. \]
Continuing in this way gives the inequality
\begin{equation} \label{E:KPdiv}
c_{p+2m,d_{p+2m}-d_p+4m} > \frac{ d_p ! p!}{(d_p-4m)! (p+2m)!} c_{p,0}
\end{equation}
for $m$ any positive integer such that $d_p \ge 4m$.

(i) Set $p:=8q$ and $m:=q$ for some positive integer $q$. Then
from Lemma \ref{L:KPdeg}(i), we have that $d_{8q} = 4q(d_0-2)+d_0$
and $d_{10q} = 5q(d_0-2)+d_0$; from Lemma \ref{L:KPcoeff}(i),
we have that $c_{8q,0} > \epsilon^{4q+1}$. Substituting these expressions
into (\ref{E:KPdiv}) completes the proof.

(ii) Set $p:=36q+1$ and $m:=3q$ for some positive integer $q$. Then
from Lemma \ref{L:KPdeg}(ii), we have that $d_{36q+1} = 12q(d_1-2)+d_1$
and $d_{42q+1} = 14q(d_1-2)+d_1$; from Lemma \ref{L:KPcoeff}(ii),
we have that $c_{36q+1,0} > \epsilon^{12q+1}$. Substituting these expressions
into (\ref{E:KPdiv}) completes the proof.

(iii) Set $p:=8q$ and $m:=q$ for some positive integer $q$. Then
from Lemma \ref{L:KPdeg}(iii), we have that $d_{8q} = 4q(d_0-2)+d_0$
and $d_{10q} = 5q(d_0-2)+d_0$; from Lemma \ref{L:KPcoeff}(iii),
we have that $c_{8q,0} > \epsilon^{2q+1}$. Substituting these expressions
into (\ref{E:KPdiv}) completes the proof.

(iv) Set $p:=36q+1$ and $m:=3q$ for some positive integer $q$. Then
from Lemma \ref{L:KPdeg}(iv), we have that $d_{36q+1} = 12q(d_1-2)+d_1$
and $d_{42q+1} = 14q(d_1-2)+d_1$; from Lemma \ref{L:KPcoeff}(iv),
we have that $c_{36q+1,0} > \epsilon^{12q+1}$. Substituting these expressions
into (\ref{E:KPdiv}) completes the proof.\qed
%which is identical to the proof for (ii)

These quantities grow factorially with $q$; thus the formal series
(\ref{E:KPnew}) diverges.

There are also holomorphic solutions in the neighbourhood of the
origin of the KP equation; for example, if the initial conditions are
linear or quadratic, then we can find exact solutions. These include
the following solutions, which are constant in $t$:
\[ u(x,y) = a_0+a_1 x + (b_0+b_1 x)y + \frac{1}{2} a_1^2 y^2
+\frac{1}{3} a_1 b_1 y^3 + \frac{1}{12} b_1^2 y^4, \]
for $u_0$ and $u_1$ linear functions of $x$ (i.e. $a_0, a_1, b_0$
and $b_1$ constants), which is holomorphic for all $x$ and $y$, and
\[ u(x,y) = \frac{(x+A)^2}{(y+k)^2} + \frac{B}{y+k} + C(y+k)^2, \]
for $u_0$ and $u_1$ quadratic in $x$ (and $k, A, B, C$ constants, $k \neq 0$),
which is holomorphic for $|y| < |k|$.

\section[]{Acknowledgements}
We are indebted to Fran\c cois Treves for
stimulating discussions in
this area.
It is a pleasure to thank the
Isaac Newton Institute, where this work was first started,
and the Australian Research Council for their
support.

\end{document}